\documentclass{iau}
\usepackage{graphicx}

\title[Seismic investigation of the $\gamma$\,Dor star KIC\,6462033: The first results of {\it Kepler} and ground-based follow up observations] 
{Seismic investigation of the $\gamma$\,Dor star KIC\,6462033: The first results of {\it Kepler} and ground-based follow up observations}

\author[Ulusoy et al.]   
{C.Ulusoy$^1$, B.Ula\c{s}$^2$, M. Damasso$^{3,4}$, A. Carbognani$^{3}$, D. Cenadelli$^{3}$, 
I. Stateva$^{5}$, I. Kh. Iliev$^{5}$, D. Dimitrov$^{5}$}

\affiliation{$^1$ College of Graduate Studies, University of South Africa, PO Box 392, UNISA 0003, Pretoria, South Africa \\email: {\tt cerenuastro@gmail.com} \\[\affilskip]
$^2$Department of Astronomy and Space Sciences, Faculty of Sciences, University of Ege, Bornova, 35100, \.{I}zmir, Turkey  \\[\affilskip]
$^3$Astronomical Observatory of the Autonomous Region of the Aosta Valley (OAVdA) Loc. Lignan 39, 11020 Nus (Aosta), Italy  \\[\affilskip]
$^4$Department of Physics and Astronomy, University of Padova, vicolo dell'Osservatorio 3, I-35122 Padova, Italy \\[\affilskip]
$^5$Institute of Astronomy with NAO, Bulgarian Academy of Sciences,blvd.Tsarigradsko chaussee 72, Sofia 1784, Bulgaria \\[\affilskip]
}

\pubyear{2013}
\volume{xxx}  
\pagerange{119--126}
\jname{Title of your IAU Symposium}
\editors{A.C. Editor, B.D. Editor \& C.E. Editor, eds.}
\begin{document}

\maketitle

\begin{abstract}
We present the first preliminary results on the analysis of ground-based time series of the $\gamma$\,Dor star KIC\,6462033 (TYC 3144-646-1, $V= 10.83$, $P=0.69686$ d) as well as {\it Kepler} photometry in order to study pulsational behaviour in this star.
\end{abstract}

\firstsection 

\begin{figure}
\begin{center}
\begin{tabular}{cc}
\includegraphics[scale=0.45]{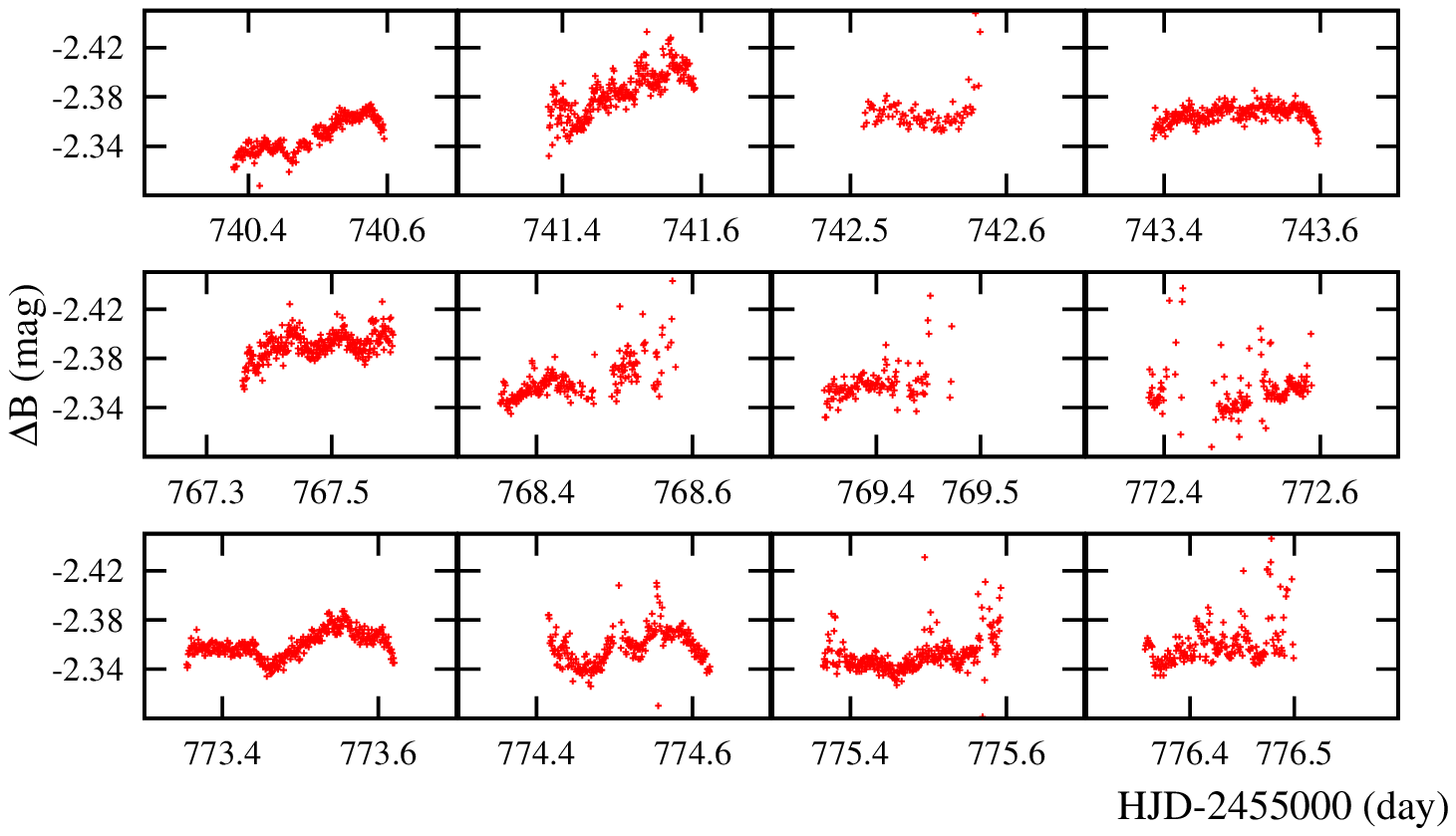} & \includegraphics[scale=0.75]{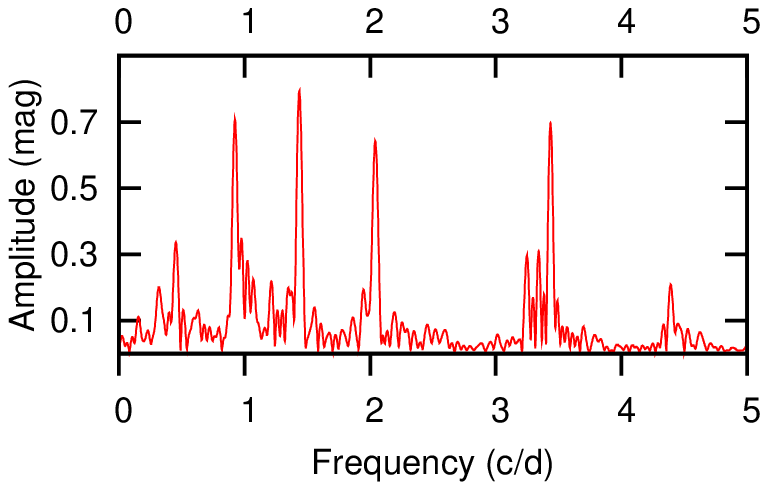}\\
\end{tabular}
\caption{{\it left:} B filter light curves of KIC\,6462033 {\it right:}Fourier spectrum of KIC\,6462033 for SC {\it Kepler} data}
\label{fig1}
\end{center}
\end{figure}
\section{Introduction}
$\gamma$\,Dor variables, which exhibit g-mode pulsations, are promising asteroseismic targets to understand their rich complexity of pulsational characteristics in detail. In order to achieve this goal, intensive and numerous multicolour and high resolution spectroscopic observations are also required, to complete space-based data aimed at the determination of their physical parameters. 

\section{Multicolour Photometry}
Ground-based observations of KIC\,6462033 were carried out between the Julian dates 740.4 and 776.5 (HJD-2455000) by using the 81-cm telescope equipped with the FLI PL3041-1-BB CCD camera in the Johnson $UBVI$ filters at the Astronomical Observatory of the Autonomous Region of the Aosta Valley (OAVdA). As can be presented in Fig.~\ref{fig1}, the B filter data are of lower accuracy due to the bad weather conditions and the data points are scattered more than expected. Because of this, more multicolour photometric data are still in need of the future work. 

\section{ {\it Kepler} Photometry}
{\it Kepler} data were used to derive the frequency content of KIC\,6462033. In this work, we analyzed only Short Cadence (SC) data with 38016 points available at the {\it Kepler} observing quarter Q3.3. Before the analysis, Simple Aperture Photometry (SAP) fluxes were cotrended and detrended following the method given by \cite{frathomp,chris12,ulu13}.
 
\section{Fourier Analysis}
The Fourier Analysis was applied to SC data by using  {\tt PERIOD04} (\cite{Lb05}). Signal-to-noise ($S/N$) threshold was adopted as 3.5 by following (\cite{Breger11}), and the periodogram of the data is also shown in Fig.~\ref{fig1}. The frequency analysis yielded four main frequencies with higher significance ($f_{1}$= 0.9242, $f_{2}$=1.4363, $f_{3}$= 2.0409 and $f_{4}$=3.4257 \,d$^{-1}$) in the range from 0.9 to 3.4 c/d. 

\section{Discussion and Conclusions}
From the analysis of ground-based and {\it Kepler} time series, we also confirmed the frequency interval [0.2,4.5]\,d$^{-1}$ given by \cite{uyt11}. Since {\it Kepler} data are not compatible with existing mode identification models, multicolour photometry is a powerful tool to test stellar theory through asteroseismic studies. In particular, some hybrid  $\gamma$\,Dor / $\delta$\,Scuti stars and other potential candidate ones have been proposed in recent works (\cite{uyt11}, \cite{BalDziem11}). For this study, we expect that near future work using high-resolution spectroscopy and multicolour photometry will clarify our results on mode identification which is a signature of the physical mechanisms ongoing in the deep stellar interior.

~\\
{\bf{Acknowledgements}}\\
\small{CU sincerely thanks the South African National Research Foundation (NRF) for
the award of NRF MULTI-WAVELENGTH ASTRONOMY RESEARCH PROGRAMME (MWGR), 
Grant No: 86563 to Prof LL Leeuw at UNISA, Reference: MWA1203150687}


\begin{thebibliography}


\bibitem[Balona \& Dziembowski 2011]{BalDziem11}
{Balona L. A. \& Dziembowski W. A.} 2011,  
\textit{MNRAS},417, 591

\bibitem[Lenz \& Breger 2005]{Lb05}
{Lenz, P.\& Breger, M.} 2005,
\textit{Commun. Asteroseismol.}, 146, 53
\bibitem[Breger et al. 2011]{Breger11}
{Breger M. et al} 2011, 
\textit{MNRAS}, 414, 1721

\bibitem[Christiansen et al. (2012)]{chris12}
{Christiansen, J. L. et al. 2012} 1999, 
\textit{Kepler Data Release 14 Notes (KSCI-19054-001)}

\bibitem[Fraquelli  \& Thompson (2012)]{frathomp}
{Fraquelli, D. \& Thompson, S. E.} 2012, 
\textit{Kepler Archive Manual (KDMC-10008-004)}

\bibitem[Ulusoy et al. (2013)]{ulu13}
{Ulusoy, C. et al.} 2013,
\textit{MNRAS}, 433,394. 

\bibitem[Uytterhoeven et al. (2011)]{uyt11}
{Uytterhoeven, K. et al.} 2011,
\textit{A\&A}, 534,125 



\end{thebibliography}
\end{document}